\documentclass[aps,prl,twocolumn,showkeys,showpacs,amsmath,amssymb]{revtex4-1}

\usepackage{graphicx}

\begin{document}

\title{Dense granular flow around a penetrating object:\\Experiments and hydrodynamic model}

\author{A. Seguin$^{1}$}
\author{Y. Bertho$^{1}$}
\author{P. Gondret$^{1}$}
\author{J. Crassous$^{2}$}

\affiliation{$^{1}$Univ Paris-Sud 11, Univ Paris 6, CNRS, Lab FAST, B\^at.~502, Campus Univ, F-91405 Orsay, France}
\affiliation{$^{2}$Univ Rennes 1, Institut de Physique de Rennes (UMR UR1-CNRS 6251), B\^{a}t.~11A,Campus de Beaulieu, F-35042 Rennes, France}


\begin{abstract}
We present in this Letter experimental results on the bidimensional flow field around a cylinder penetrating into dense granular matter together with drag force measurements. A hydrodynamic model based on extended kinetic theory for dense granular flow reproduces well the flow localization close to the cylinder and the corresponding scalings of the drag force, which is found to not depend on velocity, but linearly on the pressure and on the cylinder diameter and weakly on the grain size. Such a regime is found to be valid at a low enough ``granular" Reynolds number.
\end{abstract}

\pacs{45.70.-n, 45.50.-j}


\maketitle

{\it Introduction.}-- Describing the motion of an obstacle through granular material is the subject of recent and intensive research with applications to industrial configurations but also to biological and earth science with, example include animal locomotion in sand \cite{Maladen2009} and impact cratering \cite{Katsuragi2007}. If the motion of an object in a simple fluid is known for a long time, especially in the viscous regime at low Reynolds number where the fluid flow and the drag force are analytically known since Stokes' calculation, the motion of an object in granular matter is still an open question. Such a problem if of fundamental interest, with numerous open questions of statistical physics concerning (for instance) the solid-liquid or jamming transition~\cite{Dauchot2009}. Numerous studies have been done concerning the drag force on an object in vertical or horizontal motion in granular matter  \cite{Albert1999, Albert2000, Albert2001, Stone2004, Hill2005, Swinney2007, Zhou2007, Caballero2009}. All these studies find that the drag force does not depend on the velocity at low velocities, is proportional to the size of the object, and to its depth. As in hydrodynamics, the drag force has been shown to depend on the exact shape of the object \cite{Albert2001}, and also vertical lift forces can develop during horizontal motion \cite{Ding2011}. Flow observations of grains have been also reported in chute flow around a fixed cylinder \cite{Kellay2001, Zenit2003} and in the two-dimensional situation of a disk pulled at a constant force in a horizontal assembly of disks on a vibrated plate \cite{Dauchot2009}. Fluctuations have been observed in the force or in the velocity with some ``stick-slip" behavior in some cases \cite{Albert2000, Dauchot2009}, and the force may depend crucially on the packing volume fraction \cite{Swinney2007, Dauchot2009, Caballero2009}. In this Letter, we investigate by Particle Image Velocimetry (PIV) measurements the flow around a cylinder penetrating into a dense granular packing together with force measurements. By a continuum hydrodynamic model based on the kinetic theory extended to dense granular systems, we recover the experimental results of the shear localization close to the cylinder with the view of a ``hot" cylinder in motion in a viscosity-dependent temperature fluid, together with the good scaling for the drag force.
\begin{figure}
\includegraphics[width=0.9\linewidth]{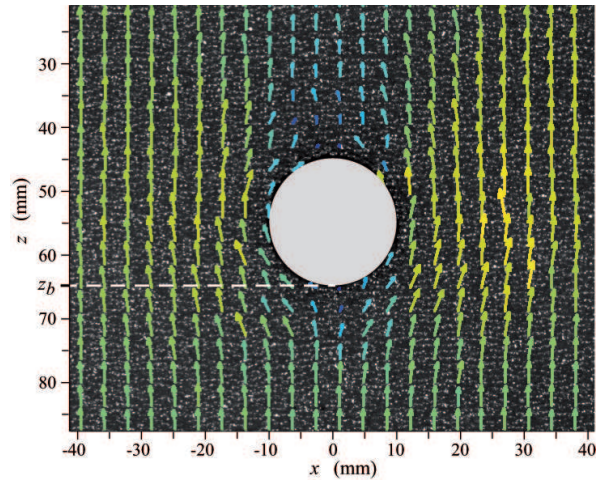}
\caption{Typical instantaneous velocity field obtained by PIV measurements for a cylinder of diameter $d=20$\,mm penetrating in glass beads of diameter $d_g=1$\,mm, at the velocity $V_0=5$\,mm s$^{-1}$ at the depth $z_b=65$\,mm.}
\label{Fig1}
\end{figure}

{\it Experiments.}--The experiments consist in a horizontal steel cylinder of diameter 10\,mm$\leq d\leq $40\,mm and length $b=40$\,mm penetrating in a rectangular box (0.1\,m in height, 0.2\,m in width and 40\,mm in thickness) filled with monodisperse millimetric glass beads of diameter 0.5\,mm$\leq d_g\leq$4\,mm and density $\rho_g = 2.5\ 10^3$\,kg m$^{-3}$. The granular medium is prepared by gently stirring the grains with a thin rod and the surface is then flatten using a straightedge. We have checked that this preparation leads to reproducible results with only small variations. The solid volume fraction is $\Phi \simeq 0.62$ characteristic of a dense granular packing and the density of the granular medium is thus $\rho = \rho_g \Phi \simeq 1.5\ 10^3$\,kg m$^{-3}$. The cylinder which is fixed and related to a force transducer by a vertical thin rod, is first above the grain surface and penetrates gradually into the granular packing as the box is raised up by a stepper motor along a vertical translation guide at a constant velocity $V_0$ ranging from 0.1 to 100\,mm s$^{-1}$. Very careful alignment is taken to prevent any blockage of the cylinder during the motion, and the force at the glass walls without grains is totally negligible. The front and rear wall of the box are in glass allowing visualization of the granular flow around the cylinder. The images taken from a fast video camera (up to 1000 images per second in the full resolution $1024\times 1024$ pixels) are analyzed by a PIV software to get the velocity field of the grains. As the camera is fixed in the laboratory frame together with the cylinder, the obtained velocity field shown in Fig.~\ref{Fig1} is the velocity of the grains in the frame of reference of the cylinder.
\begin{figure}
\includegraphics[width=\linewidth]{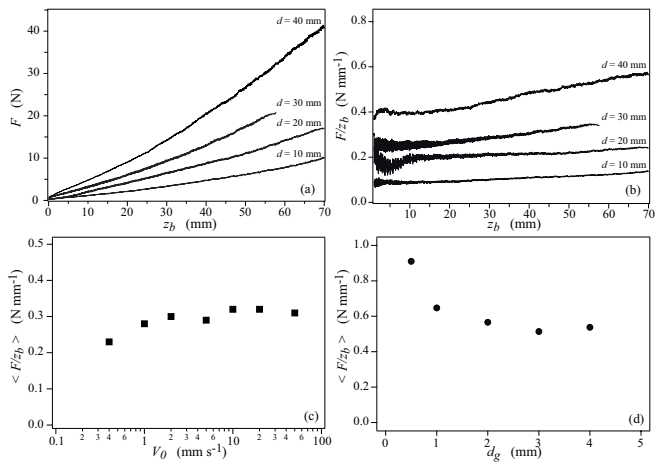}
\caption{(a)~Drag force $F$ on the cylinder as a function of the penetration depth $z_b$ for different cylinder diameters from $d=10$\,mm to $d=40$\,mm ($d_g=1$\,mm, $V_0=5$\,mm s$^{-1}$). (b)~$F/z_b$ for the same data as (a). (c)~$<F/z_b>$ as a function of $V_0$ ($d=20$\,mm, $d_g=1$\,mm) and (d) as a function of $d_g$ ($d=40$\,mm, $V_0=5$\,mm s$^{-1}$). $<F/z_b>$ is the average of $F/z_b$ over the range $d/2 \le z_b \le 70$\,mm.}
\label{Fig2}
\end{figure}

The measured drag force on the cylinder is observed to increase with the depth $z_b$ during its penetration (Fig.2a) with a ratio $F/z_b$ constant to $\pm 10 \%$ over the range $d/2 \le z_b \le 70$\,mm (Fig.2b). The force is found  proportional to the cylinder diameter (Fig.2b) and roughly independent of the velocity (Fig.2c). We find also a non-linear dependence of the force with the grain diameter: The force is about constant at large enough grain size ($d_g \gtrsim$ 1\,mm) but increases with decreasing grain size ($d_g \lesssim$ 1\,mm)(Fig.2d).
\begin{figure}
\includegraphics[width=1\linewidth]{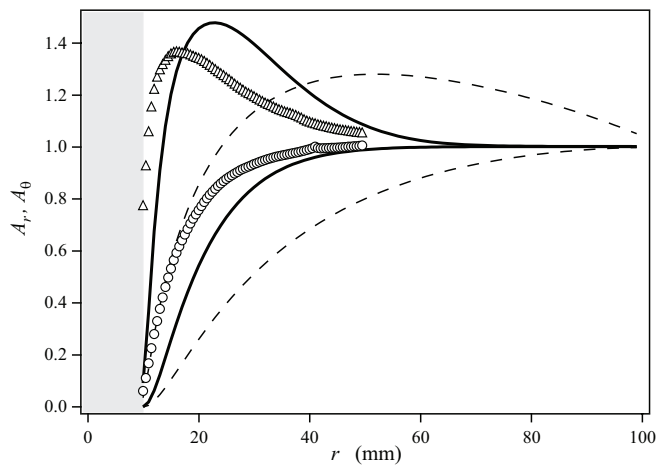}
\caption{Radial velocity functions $A_r$ and $A_\theta$ as a function of the distance $r$ from the cylinder. Experimental data ($\circ$)~$A_r$ and ($\triangle$)~$A_\theta$ from PIV measurements; (---)~Profiles obtained by using granular kinetic theory; (-- --)~Classical Newtonian profiles.}
\label{Fig3}
\end{figure}

Concerning now the grain flow field around the totally immersed cylinder, the first important and key observation is that the velocity field is stationary during the penetration process, meaning that it depends neither on the depth nor on the increasing granular pressure. The velocity field $v(x,z)$ can thus be averaged during each penetration run to extract at each point the mean local velocity $\bar{v}(x,z)$ whose time fluctuations can be related to the local granular temperature $T$, by $T = <(v-\bar{v})^2>$. Due to the geometrical configuration, we have used the most appropriate cylindrical coordinates $(r, \theta)$ where $r$ is the radial distance from the cylinder center and $\theta$ is the angle relative to the downward $z$-axis of motion (with thus $\theta = 0$ down): $\textbf{v}(r,\theta)=v_r (r,\theta)\textbf{e}_r + v_{\theta}(r,\theta)\textbf{e}_{\theta}$, with the radial and azimuthal components of the velocity $v_r$ and $v_{\theta}$. As in the classical hydrodynamics situation of a Newtonian fluid, we have checked that $v_r$ and $v_{\theta}$ can be decomposed into cosine and sine functions of $\theta$ and radial functions $A_r(r)$ and $A_{\theta}(r)$ : $v_r=-V_0 A_r(r)\cos\theta$ and $v_{\theta} = V_0 A_{\theta}(r)\sin\theta$. The radial functions $A_r(r)$ and $A_{\theta}(r)$ extracted from measurements in the azimuthal range $-\pi/2 < \theta < \pi/2$ are displayed in Fig.~\ref{Fig3} and show a strong shear localization when compared to the classical viscous Newtonian case, with exponential variations scaled by the cylinder diameter. At a distance from the cylinder surface larger than about one cylinder diameter ($r \gtrsim$ 40 mm in Fig.~\ref{Fig3}), the grain velocity vanishes. The overshoot of $A_{\theta}(r)$ expected from mass conservation in the bidimensional configuration is localized close to the cylinder and relaxes to the asymptotic value 1 (no grain flow) with an inflexion point in the present case in contrast with the long-range decay of the Newtonian case at Re = 0 (Couette-Poiseuille form).
\begin{figure}
\includegraphics[width=1\linewidth]{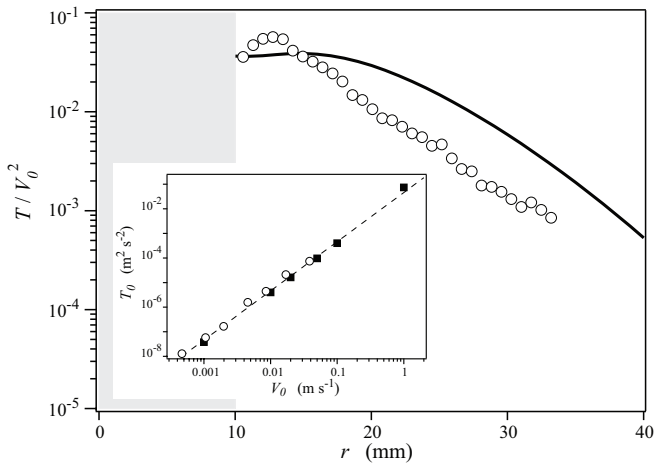}
\caption{($\circ$)~Experimental and (---)~theoretical profiles of the granular temperature $T/V_0^2$ as a function of the distance $r$ from the cylinder, for $\theta=0$. Inset: ($\circ$) experimental and ($\blacksquare$) theoretical plateau temperature $T_0$ as a function of $V_0$; (--~--)~power fit of the data of the form $T_0\propto V_0^2$.}
\label{Fig4}
\end{figure}

A typical radial profile of the granular temperature $T(r)$ along the vertical downward line ($\theta $ = 0) is displayed in Fig.~\ref{Fig4}: A domain of roughly constant temperature can be seen around the cylinder with the plateau value $T_0$ followed by an exponential decrease. The temperature profile is roughly the same for different values of $\theta$ and the plateau value $T_0$ is found to vary with the penetration velocity as $T_0 \sim V_0^2$ (see inset of Fig.~\ref{Fig4}).

{\it Hydrodynamic model.}-- The granular mean velocity and its fluctuation show important spatial variations. The strong localization of the granular flow that we observe is a rather usual feature of disordered matter where shear bands are commonly observed (see.~\cite{Schall2010} for a recent review). We show here that the observed shear bands may be understood using an hydrodynamic description. The starting points are that local momentum and energy balance equations, written as~\cite{Luding2009}:
\begin{eqnarray}
\rho {d {\bf v}\over dt}&=&\nabla \cdot \sigma ,\label{eq1}\\
\rho {d T\over dt}&=&\sigma : \kappa -\nabla \cdot {\bf q} -\varepsilon T ,
\label{eq2}
\end{eqnarray}
where $\rho$ is the effective fluid density, $\sigma$ the stress tensor, $\bf q$ the heat flux, $\kappa$ the velocity-gradient tensor, and $\varepsilon$ the temperature-loss coefficient. The stress and the heat flux are related to the velocity, temperature and pressure by phenomenological equations. The choice of these phenomenological equations to describe granular matter is still matter of debate. If momentum transfer and dissipation occur during binary collisions between grains, granular material may be treated using an inelastic gas theory~\cite{Jenkins1983}, leading to a Newtonian fluid with $\sigma=-P\mathbb{I} + 2\eta\kappa$, and Fourier's law ${\bf q}=-\lambda \nabla T$ for the heat flux. Here $P$ is the pressure, $\eta$ the viscosity, $\lambda$ the thermal conductivity and $\mathbb{I}$ the unit tensor. For simplicity, we neglect dissipation and heat transport associated with compressibility and we treat granular matter as an incompressible fluid as experimentally observed. {\it A priori}, in a dense granular flow, non-binary collisions cannot be neglected and transport coefficients from inelastic gas theory~\cite{Jenkins1983} are no longer valid. By analogy with glassy materials, modification of the viscosity divergence near jamming has been suggested~\cite{Bocquet2001}. Numerical simulations of 2D granular materials seem then to show viscosity divergence at packing fractions lower than random close packing~\cite{Garcia2006,Luding2009}. Since in our experiment, most of the shear is located in a region of high velocity fluctuations, we are not very close to random close packing, and therefore we use the Enskog expressions of the phenomenological coefficients. We obtain~\cite{Jenkins1983}:
\begin{eqnarray}
P&=&\rho_g T f_P(\Phi)\\
\eta &=& \rho_g d_g \sqrt{T} f_\eta(\Phi)\\
\lambda&=& \rho_g d_g \sqrt{T} f_\lambda(\Phi)\\
\varepsilon &=& (1-e^2) { \rho_g \sqrt{T} \over d_g} f_\varepsilon(\Phi)
\end{eqnarray}
where $f_P$, $f_\eta$, $f_\lambda$ and $f_\varepsilon$ are non-dimensional functions of the solid volume fraction $\Phi$, and $e$ is the velocity restitution coefficient. For $\Phi \gtrsim 0.5$, those functions vary with the same dependence on $\Phi$~\cite{Jenkins1983}. So we have $\eta(P,T) \simeq  \eta_0 \times (P d_g / {\sqrt T})$, $\lambda(P,T) \simeq \lambda_0 \times (P d_g / {\sqrt T})$ and $\varepsilon(P,T) \simeq \varepsilon_0 \times (P/ d_g {\sqrt T})$, with $\eta_0 \simeq 0.28$, $\lambda_0 \simeq 1.06$ and $\varepsilon_0 \simeq 0.34$~\cite{Jenkins1983} with a standard value $e\simeq 0.9$ for glass beads. Those expressions of transport and dissipation coefficients emphasize the fact that $\Phi$ is not fixed in our experiment but may vary slightly from point to point in response to pressure and granular temperature variations.

The momentum and heat equations (\ref{eq1},\ref{eq2}) with $T-P$ dependent transport coefficients are solved numerically for the stationary flow around a cylinder located at the center of a $L \times L$ square box. The momentum equation is solved using a Lattice-Bolzmann solver (BGK based D2Q9 model) with non-slip velocity conditions on the cylinder, constant pressure $P_0$ at the top of the square box, and constant upward velocity ${\bf V}_0$ on the other sides. The heat equation is solved using a finite difference scheme with the condition ${\bf e}_r \cdot \nabla T =0$ on the cylinder and $T=0$ on the sides of the square box in agreement with experimental findings. The transport coefficients are taken initially homogeneous in the box and the velocity field is first computed by solving momentum equation with these initial values. From the obtained pressure and velocity fields, the source of heat $\sigma : \kappa$ is calculated and the heat equation is then computed leading to a new temperature field $T(x,z)$. With the new corresponding fields of transport coefficients as inputs, the momentum equation and then heat equation are solved again and so on. After a few iterations, stationary temperature, pressure and velocity fields which verify (\ref{eq1},\ref{eq2}) are obtained. In order to prevent numerical instabilities, viscosity is kept in a finite range. We have checked that the stationary solution is not sensitive to the initial guess of temperature, neither to the cut-off of viscosity. We have also checked that other boundary conditions at the cylinder (partial velocity slip and Robin condition $dT/dr \propto T$ for the temperature) do not change significantly the velocity and temperature fields.

{\it Results.}--Our hydrodynamic model reproduces quite well the experimental features observed in the experiments. The radial temperature profile shows the same shape as in the experiments, with a plateau of high temperature $T_0$ close to the cylinder followed by an exponential decrease (Fig.~\ref{Fig4}). The temperature plateau is also found to be proportional to $V_0^2$ as in the experiments (see inset of Fig. 4). The computed velocity field is found close to the experimental one as shown in Fig.~\ref{Fig3} with a shear localization near the cylinder, and is far from the classical Newtonian case. This shear localization may then be interpreted as a consequence of the strong coupling between viscosity and temperature. At a given pressure the viscosity varies as $\eta \sim 1/{\sqrt T}$. So, at a given viscous shear stress $\sigma_v$, the production of heat is proportional to $\sigma_v^2 / \eta \sim \sqrt{T}$. This mechanism creates a self-lubricating layer of low viscosity near the cylinder.

The drag force on the cylinder is calculated by integrated the stresses on the disk,
taking into account both the ``pressure" term (from normal stresses) and ``viscosity" term (from shear stresses). If these two terms are equal in the Newtonian case, the pressure term is here about twice the viscosity term. The total calculated drag force is found to be independent of the velocity, proportional to the cylinder diameter and to the pressure, in agreement with the experimental observations by considering that pressure is proportional to depth. That dependance may be understood in the hydrodynamical model by considering a ``granular" Reynolds number Re = $\rho V_0 d/\eta_w$ based on the viscosity $\eta_w$ near the cylinder. We have checked that all these previous findings correspond to a low Reynolds number regime (Re $\lesssim 1$). In low Re hydrodynamics, one expects that the force scales as $\eta_w \times V_0$, and then varies here linearly with the pressure, and independently of the velocity as $\eta_w$ is proportional to the pressure and to $T_0^{-1/2}$ thus to $V_0^{-1}$.  We also found numerically the same non linear variation of the drag force with the grain size as in the experiments. Note that this variation is hard to infer simply from the set of equations. When Re $\gtrsim 1$, the velocity field no longer exhibits up/downstream symmetry, and the pressure and temperature profiles are also different from the Re $\lesssim 1$ case.

{\it Concluding remarks.}--We have investigated experimentally the penetration of a cylinder at a constant velocity inside a dense granular packing with both force and velocity field measurements, and we have modeled this problem by a continuum hydrodynamic approach. The finding of a strong shear localization close to the cylinder can be viewed by the coupling between viscosity and temperature in the problem of a self-heated cylinder. The localization of the flow near a sedimenting hot sphere has indeed already been reported in classical fluids with temperature dependant viscosity~\cite{Ansari1985}. Such a shear localization has been seen also for a sphere sedimenting in a non Newtonian fluid with shear thinning behavior~\cite{Atapattu1995}. The experimental findings of a force regime independent of velocity, proportional to the depth and to the cylinder diameter are recovered by our model based on kinetic theory adapted for dense granular systems, and have been shown to correspond to an hydrodynamic regime of low ``granular" Reynolds number. Other models exist for dense granular flows such as the one based on a local rheology with a friction coefficient $\mu(I)$ depending on a non-dimensional shear rate $I$~\cite{Pouliquen2006}. Such models may also lead to the observed flow localisation since it corresponds to visco-plastic/shear thinning behavior. Another issue to explore would be the different force value measured in the plunging and the withdrawal situations~ \cite{Hill2005, Swinney2007}, with the role played by gravity and the boundary conditions (bottom wall vs free surface). In addition, the force scaling is expected to change in a higher Reynolds regime from a ``viscous" scaling to an ``inertial" scaling which may explain the complex force terms measured in impact situations \cite{Katsuragi2007, Goldman2008, Seguin2009} where the ``granular hydrodynamic" regime change certainly from high to low Re during the penetration process. Applications to non-stationary granular flows may also been considered.

We thank J.T. Jenkins, E. Trizac and E.J. Hinch for stimulating discussions. We are grateful to F. Martinez for its contribution to the experiments and A. Aubertin for the development of the experimental setup. This work is supported by the ANR project STABINGRAM No. 2010-BLAN-0927-01.


\end{document}